\documentclass[12pt]{article}   

\usepackage{latexsym, amsfonts, amsmath, amsthm, amscd} 
\newcommand{\ot}{\otimes}

\newcommand{\boxt}{\Tox\kern -6.3pt\raise .55pt
    \hbox{$\scriptstyle{\times}$}}

\newtheorem{thm}{Theorem}[section]

\newtheorem{prop}[thm]{Proposition}
\newtheorem{cor}[thm]{Corollary}

\theoremstyle{definition}

\theoremstyle{remark}

\numberwithin{equation}{section}

\newcommand{\ovl}{\overline}

\newcommand{\C}{{\mathbb C}}
\newcommand{\Z}{{\mathbb Z}}

\newcommand{\al}{\alpha}

\newcommand{\eps}{\epsilon}

\newcommand{\la}{\lambda} 

\newcommand{\sig}{\sigma}

\newcommand{\Cnk}[1][k]{(\C^n)^{\ot #1}} 
\newcommand{\Ctwok}[1][k]{(\C^2)^{\ot #1}}
\newcommand{\Rnk}[1][k]{{\mathcal R}_{n,#1}}
\newcommand{\Rnkd}[1][d]{{\mathcal R}_{n,k,#1}}
\newcommand{\Rtwokfour}[1][k]{{\mathcal R}_{2,#1,4}}
\newcommand{\Rlast}[1][k]{{\mathcal R}_{2,#1,4}}
\newcommand{\Pnk}[1][k]{P_{n,#1}}
\newcommand{\Td}{\mathfrak T_d} \newcommand{\Ad}{\mathfrak A_d}
\newcommand{\Sd}{\mathfrak S_d}  \newcommand{\Sk}{\mathfrak S_k}
\newcommand{\fS}{\mathfrak S} 
\newcommand{\SkG}[1][G]{\Sk\,\ltimes #1}
\newcommand{\ltimes}{\vbox to 5.4pt{\leaders\vrule\vfil}\kern-4.4pt\times} 
\newcommand{\skG}[1][G]{\Sk\,\lltimes #1}
\newcommand{\lltimes}{\vbox to 3.4pt{\leaders\vrule\vfil}\kern-2.1pt\times}

\begin{document}    
 
\title { Invariant Polynomial Functions on  $k$ qudits}


\author{Jean-Luc Brylinski\thanks{Department of Mathematics, Penn State University,
  University Park 16802, \, jlb@math.psu.edu} 
  \thanks{Research supported in part  by NSF  Grant No. DMS-9803593}
   \and   Ranee Brylinski\thanks{Department of Mathematics, 
        Penn State University, University Park 16802, \, rkb@math.psu.edu}  }

%




\maketitle

\begin{abstract} We study the polynomial functions on tensor states in
$\Cnk$ which are invariant under $SU(n)^k$. We describe the space of invariant 
polynomials in terms of symmetric group representations.
For $k$ even, the smallest degree for invariant polynomials is $n$ and in  degree $n$ we 
find a natural generalization of the determinant.  For $n,d$ fixed, we describe the 
asymptotic behavior of the dimension of the space of invariants as $k\to\infty$. We study
in  detail the space of   homogeneous degree
$4$ invariant polynomial functions on $\Ctwok$.  
\end{abstract}

\section{Introduction}
\label{intro}

In quantum mechanics, a combination  of  states in Hilbert spaces
$H_1$,  .., $H_k$ leads to a  state in the tensor product Hilbert space
$H_1\ot \cdots \ot H_k$.  Such a state will be called here a tensor state.
In this paper we take $H_1=\cdots=H_k=\C^n$ where $n>1$. 
Then a tensor state is a joint state of $k$ qudits.
It would be very interesting to  classify  tensor states in $\Cnk$
up to the action of the product  $U(n)^k$ of unitary groups of local symmetries.
A natural approach to this is to study the algebra of invariant polynomials.
This approach was developed by Rains \cite{R}, by Grassl, R\"otteler and
Beth \cite{G-R-B1} \cite{G-R-B2}, by Linden and Popescu \cite{L-P} and by Coffman, 
Kundu and Wootters \cite{C-K-W}.
These authors study the ring of invariant  polynomials in the components of a tensor state in
$\Cnk$ and in their complex-conjugates. For $k$ qubits, explicit descriptions of
invariants are given in \cite{G-R-B1}, \cite{G-R-B2}, \cite{L-P} and in \cite{C-K-W}.

In this paper the symmetry group we consider is the product
 $G=SU(n)^k$ of
special unitary groups; one thinks of $G$ as the special group of local symmetries.
We study the $G$-invariant polynomial functions $Q$ on the tensor states in $\Cnk$
(we discuss in
\S \ref{sec2} how this is relevant to the description of the $G$-orbits). We consider
polynomials in the entries of a tensor state, in other words, holomorphic polynomials.

Let $\Rnkd$ be the space of homogeneous
degree $d$ polynomial functions  on  tensor states in $\Cnk$.   Let $\Rnkd^G$ be the space of  $G$-invariants in $\Rnkd$.
See \S\ref{sec2} for more discussion.  
We reduce the   problem of computing   $\Rnkd^G$  to a problem in the invariant theory of the
symmetric group  $\Sd$  (Proposition \ref{prop:RdG}). In particular, $\Rnkd^G$ is non-zero
only if
$d$ is a multiple of $n$. So the ``first"case is $d=n$; we examine this in \S\ref{sec3}.
We find that
if $k$ is odd then $\Rnkd[n]^G=0$ while if $k$ is even then $\Rnkd[n]^G$ is $1$-dimensional.
In the latter case we write  down (\S\ref{sec3}) 
explicitly the corresponding invariant polynomial $\Pnk$ in $\Rnkd[n]$; we find  $\Pnk$ 
is a natural generalization of the determinant of  a square matrix.  

For fixed $n,d$ the direct sum $\oplus_k \Rnkd$ is an associative algebra. We study the
asymptotic behavior of 
$\dim\Rnkd^G$ as
$k\to\infty$ in
\S\ref{sec4}.  In \S\ref{sec5}, we specialize to the  case of $k$-qubits, i.e. $n=2$. 
We compute the dimension of the space 
$\Rlast^G$ of degree $4$ invariants as well as
the dimension of the space of invariants in $\Rlast^G$ under the natural action of $\Sk$.  
We show that $\oplus_k\Rlast^{\skG}$ is a polynomial algebra on $2$
generators. For $k\leq 5$ we describe the representation of $\Sk$ on $\Rlast^G$. For
$k=4$ we find some interesting relations with the results on classification of tensor states in
$\Ctwok[4]$ given in \cite{B}.

We thank Markus Grassl for his useful comments on the first version of this paper.

\section{Polynomial invariants   of tensor states}
\label{sec2}

We will consider $\Cnk$ as a space of contravariant tensor states $u$. Then (once we fix a
basis of
$\C^n$) $u$  is given by  $n^k$ components $u^{p_1p_2\cdots p_k}$.
We  consider the algebra $\Rnk$ of polynomial functions on $\Cnk$.
So $\Rnk$ is the polynomial algebra $\C[x_{p_1p_2\cdots p_k}]$ in the $n^k$ coordinate 
functions  $x_{p_1p_2\cdots p_k}$. We have a natural algebra grading
$\Rnk=\oplus_{d=0}^\infty\Rnkd$ where $\Rnkd$ is the space of homogeneous degree
$d$ polynomial functions.

A  function in $\Rnkd$  amounts to a symmetric degree $d$ covariant tensor  $Q$ in
$\Cnk$. So $Q$ has  $n^{dk}$ components $Q_{i_{11}\cdots i_{dk}}$ where we think of the
indices
$i_{ab}$ as being arranged in a rectangular   array of $d$ rows and $k$ columns and  
$Q_{i_{11}\cdots i_{dk}}$ is invariant under   permutations of the  rows of the array.
Then $Q$ defines   the function
\begin{equation}
u\mapsto Q_{i_{11}\cdots i_{dk}}u^{i_{11}i_{12}\cdots i_{1k}}
u^{i_{21}i_{22}\cdots i_{2k}}\cdots u^{i_{d1}i_{d2}\cdots i_{dk}}
\end{equation}
where we used the usual Einstein summation convention.
In this way, $\Rnk$ identifies with   $S^d(\Cnk)$.

Now the group $G=SU(n)^k$ acts  on  our tensor states $u$ and tensors $Q$ as follows.  Let the
matrix
$g_{ij}$ live in the $m$-th copy of $SU(n)$ and let $g^{ij}$ be the inverse matrix. Then
$g_{ij}$ transforms 
$u^{p_1p_2\cdots p_k}$ into $g_{p_mq_m}u^{q_1q_2\cdots \cdots q_k}$ 
and  $Q_{i_{11}\cdots i_{dk}}$ into 
$Q_{j_{11}\cdots j_{dk}}g^{j_{1m}i_{1m}}g^{j_{2m}i_{2m}}\cdots g^{j_{dm}i_{dm}}$.
The identification of  $\Rnkd$ with  $S^d(\Cnk)$ is $G$-equivariant. 

We are interested in the  algebra 
$\Rnk^G=\oplus_{d=0}^\infty\Rnkd^G$ of $G$-invariants.
We view this as a first step towards studying the orbits of $G$ 
on $\Cnk$. One can first study the orbits of the complex group
$G_\C=SL(n,\C)^k$ and then decompose the $G_\C$-orbits under the $G$-action.
Note that a polynomial is $G$-invariant  
if and only if its is $G_\C$-invariant.  
The closed $G_\C$ orbits play  a special role -- they are the most degenerate orbits. 
Given any orbit $Y$, its closure  contains a unique closed orbit $Z$; 
then points in $Y$ degenerate to points in $Z$. The
$G_\C$-invariant functions  separate the closed orbits; they take the same  values on $Y$ and
on $Z$.  The set of closed orbits of $G_\C$ in $\Cnk$ has the structure of an affine complex
algebraic variety with $\Rnk^G$ as its algebra of regular functions.  Thus  a complete description
of $\Rnk^G$ would lead to a precise knowledge of the closed $G_\C$-orbits.

Our approach is thus somewhat different from that of \cite{R} \cite{G-R-B1}
\cite{G-R-B2} \cite{L-P} \cite{C-K-W} who study the  invariant  functions on
$\Cnk$ which are polynomials in the  $x_{p_1\cdots p_k}$ and in their complex conjugates;
these can also be described as the invariant  polynomial functions on
$\Cnk\oplus\ovl{\Cnk}$.

At this point it is useful to examine    the case $k=2$.
We can identify $\Cnk[2]$ with the space $M_n(\C)$ of square matrices and
then  $G=SU(n)^2$ acts on $M_n(\C)$ by $(g,h)\cdot u=guh^{-1}$.
So $\Rnkd^G$ is the space of homogeneous degree $d$ polynomial functions $Q$
of an $n$ by $n$ matrix $u$ which are bi-$SL(n,\C)$-invariant,
i.e.   $Q(guh^{-1})=Q(u)$ for $g,h\in SL(n,\C)$. Then
$Q$ is, up to scaling,  the $r$th power of the determinant  $D$ for some $r$. Hence  
$d=rn$. It follows that $\Rnk[2]^G$ is the polynomial algebra $\C[D]$.
Thus the space of closed orbits for $SL(n,\C)^2$ identifies with $\C$, where $\la$ corresponds 
to the unique closed orbit $Z_\la$  inside the set $X_\la$ of matrices of determinant  $\la$.
 For $\la\neq 0$, $Z_\la=X_\la$ while
for $\la=0$, $Z_0$  reduces to the zero matrix.

We view $S^d(\Cnk)$ as the space of invariants for the symmetric group $\Sd$ 
acting on $(\Cnk)^{\ot d}$.    So 
\begin{equation} \label{eq:2.2}
\Rnkd^G=((\Cnk)^{\ot d})^{G\times\Sd}=((\Cnk[d])^{\ot k})^{G\times\Sd}
\end{equation}
Recall the Schur decomposition
$(\C^n)^{\ot d}=\oplus_{\al}~S^{\al}(\C^n)\ot E_\al$
where $\al$ ranges over partitions of $d$ with  at most $n$ rows,
$S^\al(\C^n)$ is the irreducible covariant representation of $SU(n)$ given by the Schur
functor $S^\al$, and
$E_\al$ is the corresponding irreducible representation of  $\Sd$. 
We use the convention that 
$E_\al$ is the trivial representation if  $\al=[d]$, while 
$E_\al$ is the sign representation if  $\al=[1^d]$.
Thus we have
\begin{equation}
(\Cnk[d])^{\ot k}=\sum_{|\al_1|=\cdots=|\al_k|=d}\ S^{\al_1}(\C^n)
\ot\cdots\ot S^{\al_k}(\C^n)\ot E_{\al_1}\ot\cdots\ot E_{\al_k}
\end{equation}

Now  taking  the invariants under $G\times\Sd$ we get
\begin{equation}
\Rnkd^G=\sum_{|\al_1|=\cdots=|\al_k|=d}\ 
S^{\al_1}(\C^n)^{SU(n)}\ot\cdots\ot 
S^{\al_k}(\C^n)^{SU(n)}\ot\left(E_{\al_1}\ot\cdots\ot E_{\al_k}\right)^{\Sd}
\end{equation}
The representation $S^{\al_j}(\C^n)$, since it is  irreducible,  has no $SU(n)$-invariants except
if  $S^{\al_j}(\C^n)=\C$ is   trivial. This happens
if and only if $\al_j$ is a rectangular partition with all columns of length $n$. This proves:

\begin{prop}\label{prop:RdG} 
If $n$ does not divide  $d$, then $\Rnkd^G=0$.  If $d=nr$,  then
$\Rnkd^G$ is isomorphic to $(E_{\pi}^{\ot k})^{\Sd}$ where 
$\pi=[r^n]$. 
\end{prop}

The permutation action of $\Sk$ on $\Cnk$  induces an action
of $\Sk$ on $\Rnkd^G$.

\begin{cor} The isomorphism of Proposition \textup{\ref{prop:RdG}} intertwines the
$\Sk$-action on $\Rnkd^G$ with the action of $\Sk$ on $(E_\pi^{\ot k})^{\Sd}$
given by permuting the $k$ factors $E_\pi$.
\end{cor}

\section{The generalized determinant function}
\label{sec3}

Given $n$ and $k$, we want  to find  the smallest positive value of $d$
such that $\Rnkd^G\neq 0$.
By Proposition  \ref{prop:RdG}, the first candidate is $d=n$. 
 
\begin{cor} $\Rnkd[n]^G\neq 0$ iff  $k$   is even. In that case,  $\Rnkd[n]^G$
is one-dimensional  and consists of the multiples of  the function $\Pnk$ given by
\begin{equation}\label{eq:Pnk} 
\textstyle\Pnk(u)=\sum_{\sig_2,\cdots,\sig_k\in \fS_n}
\eps(\sig_2)\cdots\eps(\sig_k)~\prod_{h=1}^n~u^{h h_{\sig_2}\cdots h_{\sig_k}}
\end{equation}
where $h_{\sig_j}=\sig_j(h)$.
\end{cor}

\begin{proof} By Proposition  \ref{prop:RdG}, we need to compute $(E_\pi^{\ot k})^{\Sd}$.
For $d=n$, $\pi=[1^n]$ and so $E_\pi$ is the sign representation of $\fS_n$.
Then $(E_\pi^{\ot k})$ is   one-dimensional and carries the trivial representation 
if  $k$ is even, or the sign representation if  $k$ is odd.

Now for   $k$ even, we can easily compute a non-zero function $P=\Pnk$ in $\Rnkd[n]$. 
For   $S^\pi(\C^n)$    is the top exterior power $\wedge^n\C^n$. Thus $P$
is a non-zero  element of the one-dimensional subspace $(\wedge^n\C^n)^{\ot k}$ of
$(\Cnk[n])^{\ot k}$. The  tensor  components of $P$ are then given by
$P_{i_{11}\cdots i_{nk}}=\frac{1}{n!}\eps(\sig_1)\cdots\eps(\sig_k)$ if for each
$j$,  the column $i_{1j},\cdots,i_{nj}$ is a permutation $\sig_j$ 
of $1,\cdots,n$ and  $0$  otherwise. Then we get
\begin{equation}
\textstyle\label{gen_det}\Pnk(u)=\frac{1}{n!}\sum_{\sig_1,\cdots,\sig_k\in \fS_n}
\eps(\sig_1)\cdots\eps(\sig_k)~\prod_{h=1}^n~u^{h_{\sig_1}\cdots h_{\sig_k}}
\end{equation}
where $h_{\sig_i}=\sig_i(h)$. The expression   is very redundant, as each term appears
$n!$ times. We remedy this by restricting the first permutation
$\sig_1$ to be $1$. This gives (\ref{eq:Pnk}).
\end{proof}

$\Pnk$ is a \emph{generalized determinant}; $\Pnk$ is  invariant under the $\Sk$-action.
For $k=2$, (\ref{eq:Pnk}) reduces to the usual formula for the  matrix determinant.
  
Recall that the rank $s$ of a tensor state $u$ in $\Cnk$ is the smallest
integer $s$ such that $u$ can be written as $u=v_1+v_2+\cdots+v_s$,
where the
$v_i$ are decomposable tensor states
$v_i=w_{i1}\ot w_{i2}\ot\cdots\ot w_{ik}$.
There is a relation between the rank and the vanishing of $\Pnk$ as follows:

\begin{cor}
If the tensor state $u$ in $\Cnk$ has rank less than  $n$, then $\Pnk(u)=0$.
\end{cor}
It is easy to find  a tensor state $u$ of  rank $n$  such that $\Pnk(u)$ is non-zero.
For instance, $\Pnk(u)=1$ if   $u$ has all components zero except 
$u^{1\cdots 1}=\cdots=u^{n\cdots n}=1$. For $k=2$, $\Pnk(u)=0$ implies $u$ has rank
less than $n$. For bigger (even) $k$, this is false, if   $n$ is large enough. 
This happens essentially because the  rank of $u$ can be very large (at least 
$\frac{n^k}{kn-k+1}$).  Thus  $\Pnk$ gives only partial information about the rank.

\section{Asymptotics as $k\to\infty$}
\label{sec4}

Suppose we  fix  $n$ and $d$ where $d=rn$. 
Then there is a $G$-invariant  associative graded algebra structure $P\circ Q$
on the  direct sum $\oplus_k\Rnkd^G$.   Indeed, the product of tensors induces a
$(G\times\Sd)$-invariant map 
$V^{\ot k}\ot V^{\ot l}\to V^{\ot (k+l)}$  where  $V=(\C^n)^{\ot d}$.
The induced multiplication on the spaces of $(G\times\Sd)$-invariants gives the product on
$\oplus_k\Rnkd^G$, where we use the identification in (\ref{eq:2.2}). 
This multiplication corresponds, 
under the isomorphism of Proposition \ref{prop:RdG}, to the product map
$E_\pi^{\ot k}\ot E_\pi^{\ot l}\to E_\pi^{\ot(k+l)}$.
This structure is very useful. For instance, if $d=n$, then
$\Pnk[k]\circ\Pnk[l]=\frac{1}{n!}\Pnk[k+l]$. 
Thus the  determinant  $\Pnk[2]$ determines $\Pnk[2m]$ in that the $m$-fold product
$\Pnk[2]\circ\cdots\circ\Pnk[2]$ is equal to $(n!)^{-m+1}\Pnk[2m]$.

We will study the size of the algebra $\oplus_k\Rnkd^G$ by finding an asymptotic formula for     
the dimension of $\Rnkd^G$.  We do this for $r\ge 2$. Indeed for $r=1$  
we already know $\dim\,\Rnkd[n]^G$ is $1$ if $k$ is even or $0$ if $k$ is odd; we call this the 
\emph{static} case. 
The asymptotics involve  the number
\begin{equation}\label{eq:} 
p=\textstyle\dim E_\pi=d!\,\prod_{m=0}^{n-1}\frac{m!}{(m+r)!}
\end{equation} 
where $\pi=[r^n]$ as in Proposition \ref{prop:RdG}. Our formula for $p$ is immediate from the
hook formula for the dimension of an irreducible symmetric group representation.

\begin{prop}\label{prop:asymp} Assume $d=rn$ with $r\geq 2$.  Then 
$\dim\,\Rnkd^G\,\sim  c\dfrac{p^k}{d!}$
as $k\to\infty$, where  $c=1$ with one exception:   $c=4$ if $n=2,d=4$.
\end{prop}
\begin{proof} 
Let $s=\dim\,\Rnkd^G=\dim\,(E_{\pi}^{\ot k})^{\Sd}$. 
Then  $s=\frac{1}{d!}\sum_{\sig\in\Sd}\chi(\sig)^k$ where 
$\chi:\Sd\to\Z$ is the character  of  $E_\pi$. 
If $\sig$ acts trivially on $E_\pi$, then $\chi(\sig)=p$. 
If $\sig$ acts non-trivially, we claim  $|\chi(\sig)|<p$. To show this, it  suffices to show that
$\sig$ has at least two distinct eigenvalues  on $E_\pi$; this is because 
$\chi(\sig)$ is the sum of the $p$ eigenvalues of $\sig$.
Now the set $\Td$ of $\sig\in \Sd$ which act on $E_\pi$  by a scalar is a normal subgroup of
$\Sd$. So if  $d\geq 5$, then $\Td$ is $\{1\}$, the alternating group $\Ad$ or $\Sd$.
We can easily rule out the latter two possibilities, so $\Td=\{1\}$, which proves our claim.
If $d\le 4$, then (since $r>1$ and $n>1$), we have $d=4$, $n=2$ and  $\pi=[2,2]$. 
Our claim is clear here since 
$\fS_4$ acts on $E_\pi$ through the reflection representation of $\fS_3$ on $\C^2$.
 
Therefore  we have
$s=c\frac{p^k}{d!} +o(p^k)$ as $k\to\infty$ where $c$ is cardinality of the kernel of
$\Sd\to Aut\,E_\pi$. Our work in the previous paragraph computes $c$.
\end{proof}


 Proposition \ref{prop:asymp} implies that the algebra $\oplus_k \Rnkd^G$
is  far from commutative, as  it has roughly $1/N$ times the size of the tensor algebra
$\oplus_k(\C^p)^{\ot k}$.
We note however that   the $\Sk$-invariants in $\oplus_k\Rnkd^G$
form a commutative subalgebra, isomorphic to $S(E_\pi)^{\Sd}$.

\section{Quartic invariants of  $k$ qubits}
\label{sec5}

The case $n=2$ is of particular interest, as  here the qudits are qubits, and this is the case being
most discussed in quantum computation.
Here we can give some precise non-asymptotic results for the first non-static case,
namely  $\Rlast^G$. We put  $E=E_{\pi}=E_{[2,2]}$.
The proof of  Proposition \ref{prop:asymp} easily gives
\begin{cor} We have $\dim \Rtwokfour^G =\frac{1}{3}(2^{k-1}+(-1)^k)$.
\end{cor}
The first few values of $\dim \Rtwokfour^G$, starting at $k=1$, are $0,1,1,3,5,11,21,43$.
For $k=2$ and  $k=3$  the unique (up to scalar)
invariants are, respectively,  the  squared determinant  $P_{2,2}^2$ and the  Cayley
hyperdeterminant  $H_{2,3}$  (see \cite{G-K-Z}).  We note that the hyperdeterminant is
very closely related to the relative tangle of $3$ entangled qubits discussd in \cite{C-K-W}.

It would be useful to study $\Rtwokfour^G$ as a representation of $\Sk$, where $\Sk$ acts by
permuting the $k$ qubits.   The
$\Sk$-invariants in $\Rtwokfour[k]^G$ are the $(\SkG)$-invariants in $\Rtwokfour[k]$.
These $(\SkG)$-invariant polynomials are very significant as
they  separate the closed orbits of the extended symmetry group
$\SkG[SL(2,\C)^k]$ acting on $\Ctwok$. 
We can compute the dimension of the $\Sk$-invariants as follows:

\begin{prop} The dimension   of the space of $\SkG$-invariants in $\Rlast$ is   
$M_k=\left[\frac{k}{6}\right] +r_k$
where $r_k=0$ if $k\equiv 1 \bmod 6$, or $r_k=1$ otherwise. 
Furthermore the algebra $\oplus_k \Rlast^{\skG}$ is the polynomial
algebra $\C[P_{2,2}^2,H_{2,3}]$.
\end{prop}

\begin{proof} We have isomorphisms 
$\Rlast^{\skG}\simeq(E^{\ot k})^{\Sk\times\fS_3}\simeq S^k(E)^{\fS_3}$
since the representation of $\fS_4$ on $E$ factors through $\fS_3$.
Thus the algebra  $\oplus_k \Rlast^{\skG}$   identifies with $S(E)^{\fS_3}$.
Now $S(E)^{\fS_3}$
is the algebra of $\fS_3$-invariant polynomial functions on traceless $3\times 3$ diagonal 
matrices, and so is a polynomial algebra on the functions $A\mapsto Tr(A^2)$ and $A\mapsto
Tr(A^3)$. These invariants correspond (up to scaling) to $P_{2,2}^2$ and $H_{2,3}$.
The formula for the dimension follows easily.
\end{proof}

For instance, we have: $M_1=0$, $M_k=1$ for $2\leq k\leq 5$,   and  $M_6=2$. 
We remark that by replacing $S(E)^{\fS_3}$ by $\wedge(E)^{\fS_3}$,
it is easy to prove that the sign representation of $\Sk$ does not occur
in $(E^{\ot k})^{\fS_4}$ for any $k\geq 2$.

We can determine the $\Sk$-representation on $\Rlast^G$ 
for small $k$ by explicit trace computations.  For $k=2$ and $k=3$ we have the trivial
$1$-dimensional representation. For $k=4$,
we find
$\Rlast[4]^G$ is the direct sum $E_{[4]}\oplus E_{[2,2]}$.
The trivial representation $E_{[4]}$ of
$\fS_4$ is spanned by $P_{2,4}^2$, while the $2$-dimensional
representation $E=E_{[2,2]}$  is spanned
by the determinants $\Delta(ijkl)$ introduced in \cite{B}. Here $(ijkl)$ is a
permutation of $(1234)$. Given a tensor state  $u\in \Ctwok[4]$, we can view it as an element
$v$ of $\C^4\ot\C^4$, where the first (resp. second) $\C^4$ is the tensor product of the
$i$-th and $j$-th copies of $\C^2$ (resp. of the
$k$-th and $l$-th copies). Then $\Delta(ijkl)(u)$ is the determinant of $v$.
As shown in \cite{B}, the $\Delta(ijkl)$ span the  representation $E$ of
$\fS_4$. The significance of the $\Delta(ijkl)$ is that
their vanishing describes the closure of the set of tensor states in $\Ctwok[4]$
of rank $\leq 3$.   For $k=5$ the  representation  $\Rtwokfour[5]^G$ of
$\fS_5$ is $E_{[5]}\oplus E_{[2,1,1,1]}$.


\begin{thebibliography}{10}

\bibitem[B]{B} J-L. Brylinski, Algebraic measures of entanglement, quant-ph
0008031

\bibitem[C-K-W]{C-K-W} V. Coffman, J. Kundu and W. K. Wootters, 
Distributed Entanglement, preprint quant-ph/9907047

\bibitem[G-K-Z]{G-K-Z} I.M. Gelfand, M. Kapranov and A. Zelevinsky,
Discriminants, Resultants and Multidimensional Determinants, Birkh\"auser (1991)

\bibitem[G-R-B1]{G-R-B1} M. Grassl, M. R\"otteler and  T. Beth, Computing Local
Invariants of Quantum-Bit Systems, Phys. Review A 58 no. 3 (1998), 1833-1839;
also on the  Arxiv as quant-ph/9712040

\bibitem[G-R-B2]{G-R-B2} M. Grassl, M. R\"otteler and  T. Beth, Description of
Multi-Particle Entanglement through
                  Polynomial Invariants, Talk of M Grassl at the Isaac Newton Institute for
Mathematical
                  Sciences in July 1999, available on the web as
http://iaks-www.ira.uka.de/home/grassl/publications.html

\bibitem[L-P]{L-P} N. Linden and S. Popescu, On Multi-Particle Entanglement,
Forts. der Physik 46 (1998), no. 4-5, 567-578, also on the  Arxiv as quant-ph/9711016

\bibitem[R]{R} E. Rains, Polynomial Invariants of Quantum Codes,
EEE Trans. on Information Th. 46 no. 1 (2000), 54-59 




\end{thebibliography}
\end{document}